\documentclass[prb,reprint, superscriptaddress, nofootinbib, amsmath,amssymb, aps,longbibliography]{revtex4-2}

\usepackage{graphicx}
\usepackage{dcolumn}

\usepackage{bm}
\usepackage{caption}
\usepackage{float}
\usepackage{lipsum}
\usepackage{amsmath}
\usepackage{mathtools}
\usepackage{comment}
\usepackage{amssymb}
\usepackage{dsfont}
\usepackage{color}
\usepackage{tikz}
\usetikzlibrary{automata, positioning, arrows}
\usepackage{nccmath}
\usepackage{caption}
\usepackage{subcaption}
\usepackage[utf8]{inputenc}
\usepackage{babel}
\usepackage{amsmath,amsthm,amssymb}
\usepackage{graphicx}
\usepackage{physics}
\usepackage{listings}
\usepackage{enumitem}
\usepackage{microtype}
\usepackage[normalem]{ulem}
\usepackage[raiselinks=true, linktoc=all,hidelinks]{hyperref}
\usepackage[capitalise]{cleveref}
\newcolumntype{L}{>{$}l<{$}}

\global\long\def\av#1{\left\langle #1 \right\rangle }

\allowdisplaybreaks[1]

\begin{document}

\preprint{APS/123-QED}

\title{Fluctuating field series: towards calculations of correlated systems with high accuracy}

\begin{abstract}
  We introduce regular series expansion for weakly- and moderately-correlated fermionic systems, based on Fluctuating Local Field approach. The method relies on the explicit account of leading fluctuating mode(s) and is therefore suitable for medium-sized lattices. It can be also used as a solver for the cluster approximations for infinite-size lattices. We introduce classical fluctuating field coupled to fermionic collective mode(s). This way, fluctuations in selected modes are treated in a non-perturbative way. Other degrees of freedom are accounted for the diagram expansion performed at each value of the fluctuating field. The method is benchmarked for the $U/t=1$ and $U/t=2$ Hubbard lattices at half-filling. Results for susceptibility in the antiferromagnetic channel along with the spectral function are compared with the numerically exact quantum Monte Carlo data. Calculations up to the third order of the series expansion are performed and show a uniform convergence to the reference result for the susceptibility. This convergence is observed well below the DMFT Ne\'el temperature and makes our practically simple method applicable in a much wider temperature range than DMFT-based diagrammatic schemes. 
\end{abstract}
\author{Ya.~S. Lyakhova}
\affiliation{Russian Quantum Center, Moscow 121205, Russia}
\affiliation{National Research Nuclear University MEPhI, Moscow 115409, Russia}
\author{S.~D. Semenov}
\affiliation{Russian Quantum Center, Moscow 121205, Russia}
\affiliation{Moscow Institute of Physics and Technology, Dolgoprudny, 141701, Russia}
\author{A.~I. Lichtenstein}
\affiliation{Institut f{\"u}r Theoretische Physik, Universit{\"a}t Hamburg, Notkestra{\ss}e 9, 22607 Hamburg, Germany}
\author{A.~N. Rubtsov}\email{ar@rqc.ru}
\affiliation{Russian Quantum Center, Moscow 121205, Russia}
\affiliation{Lomonosov Moscow State University, Moscow 119991, Russia}

\date{\today}

\maketitle

\section{Introduction}

An accurate description of fluctuation phenomena in correlated materials remains one of challenging problems in condensed matter theory, in particular when it comes to correlated systems \cite{Sachdev2022,Levin2024}. Despite the clear importance of numerical modelling of the phase transition phenomena, modern theoretical tools do not always provide their predictive description. A two-dimensional Hubbard model is a particularly known example of a system being hard for a quantitatively accurate modelling. This model is of a great importance because it is believed to describe the electronic properties of a number of intriguing systems: high-Tc cuprates \cite{white2024}, twisted bilayer graphene \cite{Bernevig2022} or ultracold atoms \cite{Bloch2023}. 

The challenges encountered when modeling the Hubbard systems arise from at least three sources. First, when Hubbard on-site repulsion $U$ is compatible with bandwidth, the Hubbard model exhibits strong correlation effects, mostly local on sites and short-ranged.  Second, collective fluctuations of different nature may be present in the system. The role of fluctuation phenomena is greatly enhanced in a two-dimensional or layered systems, compared to the bulk materials. Third, a large number of possible ordered phases with very similar free energy have to be considered for the doped Hubbard model. One can mention antiferromagnetically-ordered, striped, superconducting, spin wave, charge density wave and other phases. Theoretical description of all these mutually influencing phenomena requires highly accurate numerical methods capable of taking into account different energy and length scales. Various approaches, mostly based on the pure mean-field or mean-field-like schemes, aim to establish reliable information about the phase diagram of the Hubbard system. One of frequently used approaches here is the perturbative extension. Such a way demands a non-trivial calculational schemes, as the Hubbard model is non-perturbative even at weak-coupling limit \cite{PhysRevB.65.081105}. Low-order mean-field description, though tempting for comparatively small $U$, fails to catch the ordered state physics of Hubbard lattice even qualitatively. In this case parental method is usually extended. GW method built on the basis of the Density Functional Theory \cite{RevModPhys.87.897, PhysRevB.109.134424}, Dual Fermions \cite{PhysRevB.79.045133, van_Loon_2021}, TRILEX \cite{PhysRevB.96.104504} and D-TRILEX \cite{PhysRevB.103.245123} represent different implementations of this approach. Other ways to increase the accuracy of scheme is to use hybrid methods or various cluster extensions \cite{maier2005}. A birds eye look to the existing schemes is given in the review \cite{UFN_LAR}. A particular family of diagram methods constructed around the Dynamical Mean Field Theory (DMFT) is surveyed in \cite{DFrev}. State of the art results for the Hubbard model obtained by different computational schemes are outlined in \cite{Hubbard_comp}.

The paper \cite{PhysRevX.11.011058} gives an important survey of the results obtained by different extensions of the DMFT for a particular case of the half-filled Hubbard model with $U=2 t$, where $t$ is the hoping amplitude between the neighboring sites. This case is much simpler than a generic Hubbard model with arbitrary parameters. First, half-filling ensures that the model shows fluctuations and ordering solely in the antiferromagnetic (AF) channel, so one should not deal with an interplay between different modes and candidate states. Second, the system at $U=2 t$ is moderately correlated. It could be expected that its description is simpler than the strongly correlated case of $U \gtrsim 4  t$. The results can be summarized as follows. The parental approximation, DMFT, predicts a N\'eel transition at the inverse temperature $\beta \approx 12$. This is an artifact, because Mermin-Wagner theorem forbids formation of the antiferromagnetic ordering in a planar system at finite temperature. The extended schemes indeed get rid of this transition. Even more, a very reasonable accuracy is seen when calculating both single-particle (the self-energy) and two-particle (the AF susceptibility and correlation length) properties, but only for $\beta  \lesssim 15$. We conclude that the domain of applicability of the known DMFT extensions just slightly exceeds the thermal range defined by the N\'eel temperature of the parental approximation.

Recently, there was presented a novel approach to the description of correlated systems, called Fluctuating Local Field (FLF) \cite{Rubtsov2018}. This method starts with the exact transformation of the partition function, introducing an ensemble of systems at different values of artificial fluctuating field. In practice FLF is based on a parental scheme, e.g. stationary mean field theory \cite{PhysRevB.105.035118,JSNM-LR,Stepanov1,Stepanov2} or DMFT \cite{RSL-2020}. The scheme has proved its effectiveness for weakly correlated systems, getting rid of the above-mentioned artifacts and providing good accuracy at low calculational cost.

Here, we construct a perturbation theory based on the FLF approach. Operating in a methodological spirit, we employ a simple example of the 2D Hubbard system at half-filling to illustrate our approach. Use of a well-described model with known results for a range of observables allows us to validate the reliability of the presented method. For the sake of clarity and simplicity we operate under assumption that there is a single AF fluctuating mode, that is, the system under consideration is assumed not large in size. Moreover, we limit ourselves with weakly- and moderately-correlated systems, which allows us to work with the formal expansion  w.r.t. the order of Hubbard interaction $U$. Though the two-dimensional Hubbard model itself does not have a small parameter at weak-coupling \cite{PhysRevB.65.081105}, it was noticed in \cite{JSNM-LR} that the FLF transformation leads to appearance of the effective long-range interaction in this system which competes with the on-site repulsion and thus effectively form a new small parameter. The main result of the present paper is that the FLF expansion converges regularly within a wide temperature range, much larger than the domain of applicability of usual weak-coupling expansion. This opens a pathway to future extension of the method to strongly correlated systems by constructing the FLF series on top of the DMFT result. 

We start with the formulation of FLF formalism in Sec. \ref{sec:system_formalism}, and derivation of the perturbation series for the FLF free energy in Sec. \ref{sec:Perturbation}. It allows us to calculate the single-particle and two-particles observables of the Hubbard system in Sec. \ref{sec:single} and \ref{sec:susceptibility}. Limiting ourselves with the single-mode assumption, we extend our approach to large systems by constructing the FLF cluster extension in the spirit of Cluster DMFT (CDMFT) in Sec. \ref{sec:cluster}. Having conducted all the mentioned steps for weakly correlated regime $U = 1$, we move to more detailed consideration of possible divergences and examine the $U=2$ case, which is more illustrative here. We provide a regularization procedure, smoothing the disconvergence of the perturbation series in moderately correlated regime in Sec. \ref{sec:Landau}. Comparison of the results obtained within the FLF perturbation approach with the DMFT and numerically exact Quantum Monte-Carlo (QMC)~\cite{DQMC} data demonstrates high accuracy for Green's function, local spectral density function and the susceptibility. We discuss the results of our work and provide the outlook in Sec. \ref{sec:conclusion}.

\section{System and formalism}
\label{sec:system_formalism}
We consider the two-dimensional Hubbard system with $N$ sites at half-filling. It can be described by the following Hamiltonian:

\begin{align}
\label{eq:Hubbard}
    H = & -t\sum_{\langle jj^{\prime}\rangle,\sigma} \left(c^\dag_{j,\sigma}c_{j^{\prime},\sigma} +  c^\dag_{j^{\prime},\sigma}c_{j,\sigma}\right) + \\
    \nonumber
    &+ U\sum_{j}\left(n_{j\uparrow} - \frac{1}{2}\right)\left(n_{j\downarrow} - \frac{1}{2}\right) - \sum_{j} h^z \left(n_{j\uparrow} -n_{j\downarrow}\right).
\end{align}

Here, the first term describes the nearest-neighbour pair $\langle jj^{\prime}\rangle$ hoppings with the amplitude $t$, the second term describes the on-site Coulomb repulsion, and the third term couples the magnetic moment of the lattice to the local external magnetic field $\boldsymbol{h}_j$ along the $z$-axis. Moving forward, we will discuss only the case of antiferromagnetically (AF) ordered field $\boldsymbol{h}_j = \boldsymbol{h} e^{i \boldsymbol{K} \boldsymbol{r}_j }$ with $\boldsymbol{K} = \left( \pi, \pi \right)$ and the real space vector $\boldsymbol{r}_j$.

The partition function of the Hubbard system can be expressed symbolically in the path integral formalism as:
\begin{equation}
\label{eq:Z_exact}
    Z = \int e^{- \left[\mathcal{S}^0_h + \mathcal{S}_U \right]}\mathcal{D}[c^{\dagger}, c],
\end{equation}
where $\mathcal{S}_h^0$ is the action of non-interacting system in external field $\boldsymbol{h}$, and $\mathcal{S}_U$ is the Hubbard interaction term action. We conclude the full action:
\begin{equation}
\label{eq:Full action intial Hubbard}
    \mathcal{S}_h[c^\dag, c] = \underbrace{-c^{\dagger} G_0^{-1} c -  \beta N h^ls^l}_\text{$\mathcal{S}^0_h$} +\underbrace{U\theta}_\text{$\mathcal{S}_U$}, 
\end{equation}
where $G_0$ is a bare Green's function, $\beta$ is reverse temperature, and

\begin{align}
\nonumber
    &\mathcal{S}_U = U\theta \\
    &=U\int_0^\beta\sum_j\left(c_{j\uparrow}^\dag(\tau)c_{j\uparrow}(\tau)-\frac{1}{2}\right)\left(c_{j\downarrow}^\dag(\tau)c_{j\downarrow}(\tau) - \frac{1}{2}\right)d\tau
\end{align}
is an abbreviation of the on-site interaction in the action formalism. $h^ls^l$ represents the scalar product between external field $h^l$ and collective AF static spin variable:
\begin{equation}
\label{eq:total spin}
s^l = \frac{1}{\beta N}\sum_{j \sigma \sigma^{\prime}} \int_0^\beta c_{j \sigma \tau}^{\dagger} \sigma^l_{\sigma \sigma^{\prime}} c_{j \sigma^{\prime} \tau} e^{i\boldsymbol{K}\boldsymbol{r}_j} d \tau,
\end{equation}
where $\sigma^l_{\sigma\sigma'}$ is the $l-$Pauli matrix, and the integration is w.r.t. to the imaginary time $\tau$.

\subsection{Fluctuating local field formalism}
The Fluctuating Local Field (FLF) formalism can be developed as follows: one should define the instability channel and the corresponding order parameter; couple the artificially introduced field to this order parameter; and let this field fluctuate in the controlled way. Formally, the FLF approach starts with the exact transformation of the partition function:
\begin{equation}
\label{eq:FLF_transf}
    Z = \left( \frac{\beta N}{2 \pi \lambda}\right)^{3/2} \iint e^{- \mathcal{S}_h  - \frac{\beta N}{2\lambda}\left( \boldsymbol\nu - \boldsymbol{h} - \lambda \boldsymbol{s}\right)^2} \mathcal{D} \left[ c^{\dagger}, c\right] d^3 \nu.
\end{equation} 
Here we introduced the fluctuating field $\boldsymbol\nu$, which is a classical field coupled to the AF spin variable $\boldsymbol{s}$, as we consider the AF instability channel. Parameter $\lambda$ can be considered as a gauge parameter of transformation (\ref{eq:FLF_transf}).
We will use this freedom of choice later. In what follows, we will omit the pre-integral coefficient for the sake of shortness, as it does not affect the physics.

Expanding the brackets in (\ref{eq:FLF_transf}) leads to the following expression: 
\begin{align}
\label{eq:FLF_transf simplified}
\nonumber
    Z &= \iint e^{- \mathcal{S}_{\nu}^{\prime}} e^{- \frac{\beta N}{2 \lambda}\left( \boldsymbol\nu - \boldsymbol{h} \right)^2} \mathcal{D} \left[ c^{\dagger}, c\right] d^3 \nu =\\
    &=\int \mathcal{Z}_\nu e^{- \frac{\beta N}{2 \lambda}\left( \boldsymbol\nu - \boldsymbol{h} \right)^2}d^3 \nu,
\end{align} 
where the new action $\mathcal{S}_{\nu}^{\prime}$ consists of two parts:
\begin{equation}
\label{eq:FLF action}
    \mathcal{S}_{\nu}^{\prime}[c^\dag, c] = \underbrace{-c^{\dagger} G_0^{-1} c - \beta N \nu^l s^l}_\text{$\mathcal{S}^0_{\nu}$} +\underbrace{U\theta + \frac{\beta N\lambda}{2} s^ls^l}_\text{$\mathcal{S}_{int}^{\prime}$}.
\end{equation}
Here $\mathcal{S}^0_{\nu}$ is a bare fermionic action polarized by the external field $\boldsymbol\nu$, and $\mathcal{Z}_\nu$ is the partition function of the system with action $\mathcal{S}'_\nu$ at fixed $\boldsymbol\nu$. The new effective interaction action $\mathcal{S}_{int}^{\prime}$ now involves two components, the Hubbard on-site interaction $U\theta$, which tends to the AF order, and the effective long-range ferromagnetic interaction $\frac{\beta N\lambda}{2} s^ls^l$. This spin-spin interaction weakens the AF fluctuations, thus indeed allowing us to treat them as approximately gaussian. Full range of fluctuations is then recovered via the integration over $\boldsymbol\nu$.

Eq. (\ref{eq:FLF_transf simplified}) describes the partition function of the Hubbard system $Z$ as the average over the gaussian ensemble of systems at fixed value of fluctuating field $\boldsymbol{\nu}$ with the partition function $\mathcal{Z}_\nu$. Practical implementation of this approach demands that we approximate $\mathcal{Z}_\nu$ in some way. The simple and still effective method is to use the mean-field-like approximation for this. Such an approach was applied for this task for instance in \cite{PhysRevB.105.035118} and \cite{RSL-2020}, where the authors developed the FLF method on the basis of stationary and dynamical mean field theories respectively. In this work we propose to exploit the effective small parameter $\mathcal{S}'_{int}$ and obtain $\mathcal{Z}_\nu$ perturbatively. Once this is done we will be able to calculate physical observables $\mathcal{O}$ as an average over the FLF ensemble
\begin{equation}
\label{eq:FLF_average}
    \av{\mathcal{O}}_{FLF} = Z^{-1}\int \mathcal{O}(\boldsymbol{\nu}) \mathcal{Z}_\nu e^{-\frac{\beta N}{2\lambda}\left(\boldsymbol{\nu}-\boldsymbol{h}\right)^2}d^3\nu.
\end{equation}
In the particular case of $\boldsymbol{h}\to0$ and for observables $\tilde{\mathcal{O}}$, which are spherically symmetric w.r.t. $\boldsymbol{\nu}$, this expression can be reduced to 1D-integrals
\begin{equation}
\label{eq:FLF_average_sym}
    \av{\tilde{\mathcal{O}}}_{FLF} = \frac{4\pi\int_0^{+\infty} \tilde{\mathcal{O}}(\nu) \mathcal{Z}_\nu e^{-\frac{\beta N}{2\lambda}\nu^2}\nu^2d\nu}{4\pi\int_0^{+\infty} \mathcal{Z}_\nu e^{-\frac{\beta N}{2\lambda}\nu^2}\nu^2d\nu}.
\end{equation}

\section{Perturbation theory series}
\label{sec:Perturbation}
In this section we develop the perturbation theory series, assuming that $\mathcal{S}'_{int}$ is a small parameter. As we will see later (Eq. (\ref{eq:lambda})), the parameter $\lambda$ is proportional to the Hubbard interaction constant $U$. Hence, the perturbation expansion is constructed w.r.t. orders of $U$. We will derive the corrections to the quantities of non-interacting zeroth approximation at first. Once this is done, we will find the conditions ensuring the validity of this approach. We start with the expansion of the partition function $\mathcal{Z}_\nu$:
\begin{widetext}
\begin{equation}
\label{eq: Taylor expansion}
\mathcal{Z}_{\nu} = \int \left[ 1 - \left(U\theta + \frac{\beta N \lambda}{2}s^ls^l\right)  + \frac{1}{2}\left(U^2\theta\theta + \beta NU\lambda \theta s^ls^l + \frac{(\beta N\lambda)^2}{4}s^ls^ls^{l'}s^{l'}\right)+\dots \right] e^{-\mathcal{S}_0^{\nu}} \mathcal{D} \left[ c^{\dagger}, c\right].
\end{equation}
\end{widetext}
Here and in the following, summation over the repeated indices is assumed. Now we may note that
\begin{equation}
\label{eq:derivative_trick}
    s^ms^m e^{-\mathcal{S}^\nu_0} = \frac{1}{(\beta N)^2}\frac{\partial^2}{\partial \nu^m \partial \nu^m}e^{-\mathcal{S}^\nu_0}=\frac{1}{(\beta N)^2}\Delta_\nu e^{-\mathcal{S}^\nu_0}.
\end{equation}
Let us denote $\mathcal{Z}^0_\nu = \int e^{-\mathcal{S}_\nu^0}\mathcal{D}\left[c^\dag, c\right]$, which is the partition function of non-interacting system at fixed value of $\boldsymbol\nu$ (compare with (\ref{eq:FLF_transf simplified})). The average over this ensemble reads
\begin{equation}
    \langle \dots \rangle_0 = \left(\mathcal{Z}^0_\nu\right)^{-1}\int \dots e^{-\mathcal{S}_\nu^0} \mathcal{D}\left[c^\dag, c\right].
\end{equation}
It allows us to rewrite (\ref{eq: Taylor expansion}) in the form
\begin{widetext}
\begin{equation}
\label{eq: Taylor expansion Delta}
\mathcal{Z}_{\nu} = \left[1 - \left(U\langle \theta\rangle_0 + \frac{\lambda}{2\beta N }\Delta_\nu\right)  + \frac{1}{2}\left(U^2\langle \theta\theta\rangle_0 + \frac{U\lambda}{\beta N} \Delta_\nu \langle \theta\rangle_0 + \frac{\lambda^2}{(2\beta N)^2}\Delta_\nu\Delta_\nu\right)+\dots\right]\mathcal{Z}^0_\nu.
\end{equation}
\end{widetext}

It is useful to express this result in terms of free energy. Taking the definition $\mathcal{Z}_{\nu} = \text{exp}\left(-\beta N F_\nu\right)$, we can derive the perturbation series
\begin{equation}
\label{eq:F_series}
    F_\nu \approx F_\nu^{(n)} = F^{0}_\nu + \delta F^1_\nu + \delta F^2_\nu +\dots + \delta F^n_\nu,
\end{equation}
where $F^{0}_\nu$ is the free energy of non-interacting system, and $F_\nu^{(n)}$ is the $n$th order approximation of exact free energy, including up to the $n$th order corrections $\delta F^n_\nu$. In the particular case of negligible external magnetic field $\boldsymbol{h}\to 0$ we have:
\begin{equation}
\label{eq:F_cor_1}
        \delta F^1_\nu = U\Theta_1 + \frac{\lambda}{2}(F^0_\nu)^{'2} - \frac{\lambda}{2\beta N}\left[\frac{2}{\nu}(F^0_\nu)' + (F^0_\nu)^{''}\right],
\end{equation}
\begin{widetext}
    \begin{alignat}{3}
    \nonumber
    \delta F^2_\nu = -& \frac{U^2}{2} \Theta_2 + \lambda U (F^0_\nu)' \Theta_1' - \frac{\lambda U}{2\beta N}\left[\frac{2}{\nu}\Theta_1' + \Theta_1^{''}\right] &&\\
    \nonumber
    &+ \frac{\lambda^2}{2}(F^0_\nu)^{'2}(F^0_\nu)^{''} + \frac{\lambda^2}{8\beta N}\left[\frac{4}{\nu^2}(F^0_\nu)^{'2} -2(F^0_\nu)^{''2}\right. &&\left.- \frac{8}{\nu}(F^0_\nu)'(F^0_\nu)^{''}-4(F^0_\nu)'(F^0_\nu)^{'''}\right] \\
    & &&+ \frac{\lambda^2}{8(\beta N)^2}\left[\frac{4}{\nu}(F^0_\nu)^{'''} +(F^0_\nu)^{(IV)}\right].
    \end{alignat}
\end{widetext}
Here we used the notation $\beta N\Theta_1 = \langle \theta\rangle_0$, $\beta N \Theta_2 = \langle \theta\theta \rangle_0 - \langle \theta\rangle_0 \langle \theta\rangle_0$. The derivatives are taken w.r.t. the magnitude of fluctuating field $\boldsymbol{\nu}$. The terms are grouped according to their extensity/intensity, i.e. their dependence on $\beta N$. See Appendix for the derivation of the third order corrections to free energy.

\begin{figure}[H]
\begin{minipage}[h]{\linewidth}
\center{\includegraphics[width=1\linewidth]{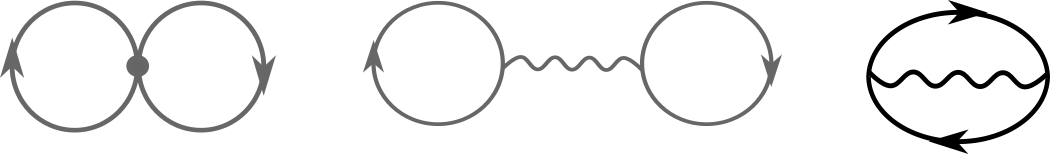}} a) \\
\end{minipage}
\vfill
\begin{minipage}[h]{\linewidth}
\center{\includegraphics[width=1\linewidth]{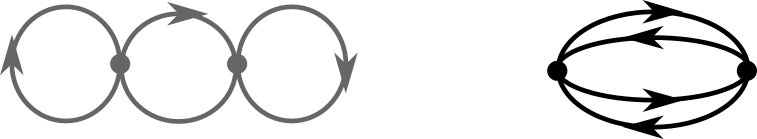}} b) \\
\end{minipage}
\vfill
\begin{minipage}[h]{\linewidth}
\center{\includegraphics[width=1\linewidth]{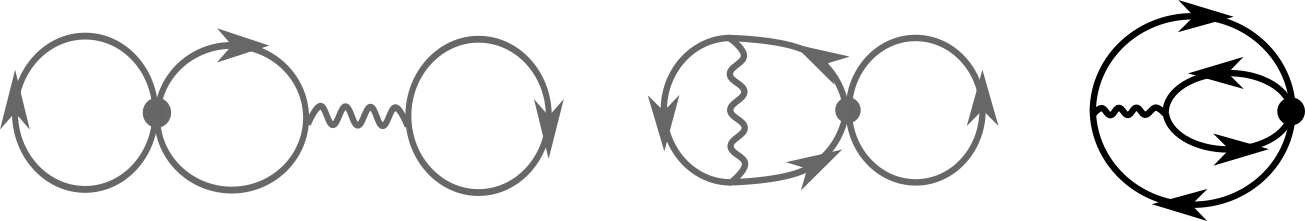}} c) \\
\end{minipage}
\vfill
\begin{minipage}[h]{\linewidth}
\center{\includegraphics[width=1\linewidth]{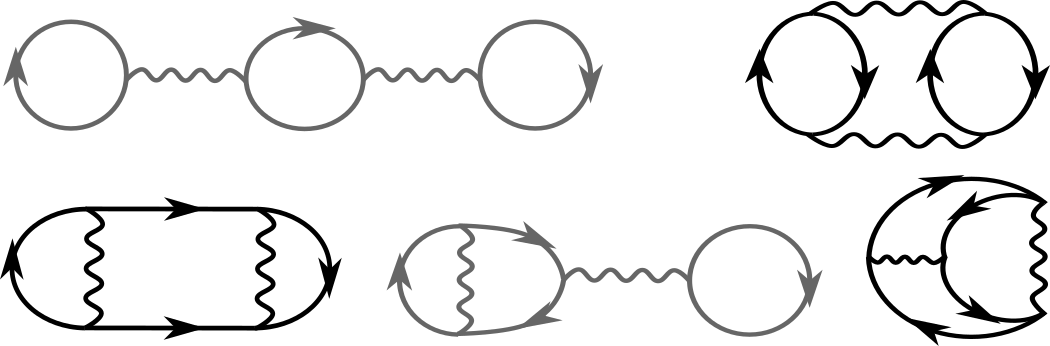}} d) \\
\end{minipage}
\caption{First order (a) and second order (b, c, d) diagrams. Bold dots denote local Hubbard interaction $U$ and curved lines denote effective long-range FLF interaction $\lambda$. Solid lines with arrows mean free fermionic propagators. Gray diagrams cancel each other at $\lambda=U/2$.}
\label{fig:First_Order_Diag}
\end{figure}

Note that the trick with derivatives in (\ref{eq:derivative_trick}) is not the only way to derive perturbation series. The same result can be obtained using the Feynman diagrams technique. According to the linked-cluster theorem \cite{coleman2015}, the free energy is determined by only connected diagrams, which simplifies the task significantly. For instance, there are only three connected diagrams of the first order (see Fig. \ref{fig:First_Order_Diag}a). The local one, which is proportional to $U$ and is denoted with a bold dot, represents the on-site Hubbard term. And the effective long-range interaction proportional to $\lambda$ is expressed through diagrams with curved lines. The first two diagrams correspond to the first two terms in (\ref{eq:F_cor_1}). They are both built of local propagators, which means they represent extensive quantities. Similar reasoning can be applied to the higher order corrections. The second order diagrams are shown in Figs. \ref{fig:First_Order_Diag}b-d. For the third order diagrams see Appendix.

Up to this point we conducted the derivation assuming that $\mathcal{S}'_{int} \to 0$. Now we can figure out whether this assumption is valid. To do this we consider the first correction to the free energy (\ref{eq:F_cor_1}). We write down explicitly what the derivatives of the free energy $F_\nu^0$ mean:
\begin{align}
\nonumber
    F_\nu^0 &= -\frac{1}{\beta N}\ln \int e^{c^\dag G_0^{-1}c-\beta N\nu^lS^l}\mathcal{D}\left[c^\dag,c\right],\\
    (F_\nu^0)' &= \av{s^z}_0, \quad (F_\nu^0)^{''} = \beta N\left(\av{s^z}_0^2 - \av{s^zs^z}_0\right).
\end{align}
Consider now the first two terms of (\ref{eq:F_cor_1}). One can see that
\begin{equation}
\label{eq:lambda}
    \av{\theta}_0 \equiv \av{\left(n_{j\uparrow}-\frac{1}{2}\right)\left(n_{j\downarrow}-\frac{1}{2}\right)}_0 = -\frac{1}{2}\av{s^z}_0^2,
\end{equation}
and thus, these two terms cancel each other at $\lambda_0 = U/2$. The last, intensive term of $\delta F_\nu^1$ corresponds to the Hartree-Fock approximation. The same result for $\lambda$ can be obtained if we consider the MF as the saddle-point approximation of the FLF ensemble (\ref{eq:FLF_average}). See \cite{RSL-2020} for details. We keep $\lambda_0=U/2$ for all of the following numerical calculations.

At this point it is worthwhile to mark the domain of applicability of our scheme. The hierarchy of diagrams in powers of $(\beta N)^{-1}$ indicates the limits on the lattice size and the temperature. Namely, the FLF perturbative approach is applicable for medium-sized lattices at moderate temperatures, such that it possesses collective excitations and at the same time the physics is still defined by a single collective mode.

\begin{figure}
    \centering
    \includegraphics[width=0.9\linewidth]{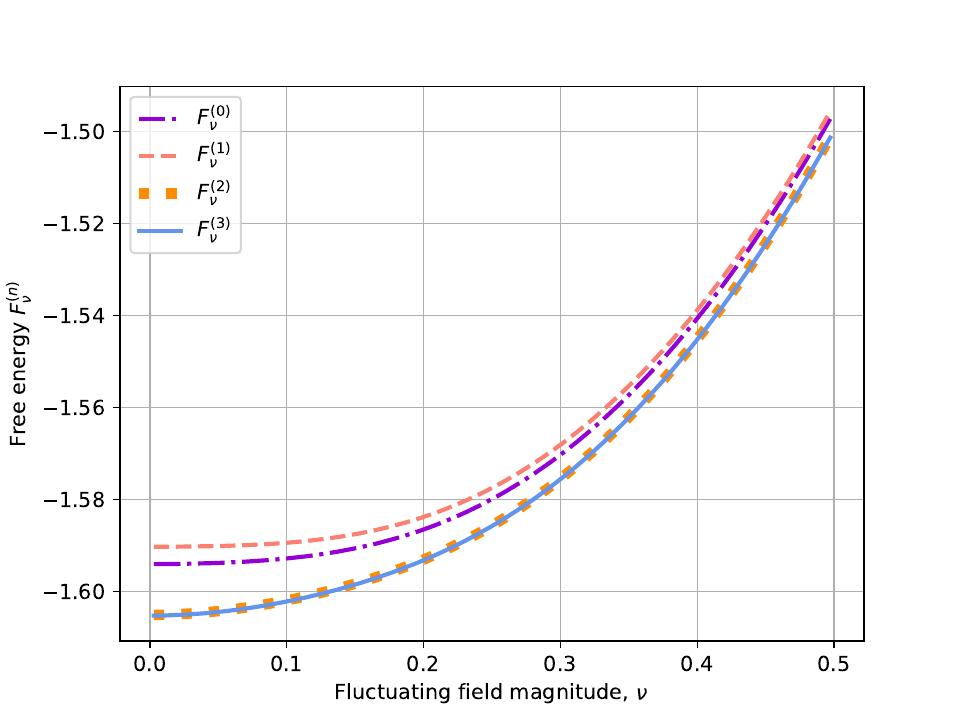}
    \caption{FLF free energy approximation $F^{(n)}_\nu$ for $n = 0,\,1,\,2,\,3$ as a function of the fluctuating field magnitude $\nu$. The parameters of the Hubbard system are $U/t = 1$, $\beta = 10$, the size of the lattice is $6 \times 6$.}
    \label{fig:F_series}
\end{figure}

Next, we will move to the practical implementation of our method. In this work we limited ourselves with the 3rd order corrections to keep the calculations simple and fast, as is provided by the FLF. Specifically, the 4th order diagrams include four-leg vertices, which requires comparatively heavy calculations. Fortunately, the 3rd order already provides us with high accuracy results. The numerical results for the free energy corrections up to the 3rd order are presented in the Fig. \ref{fig:F_series}. One can see that the 3rd order approximation is very close to the 2nd order one, demonstrating convergence. Overall, it leads us to the conclusion that introduction of the effective long-range interaction indeed allows us to develop the convergent perturbation series. Finally, it enables us to calculate the physical observables (\ref{eq:FLF_average}) and (\ref{eq:FLF_average_sym}) up to the needed order of accuracy $n$ using that
\begin{equation}
\label{eq:Z_approx}
    \mathcal{Z}_\nu \approx \mathcal{Z}^{(n)}_\nu = e^{-\beta N F^{(n)}_\nu}.
\end{equation}
In what follows we will refer to the $n$-order approximation of the FLF ensemble as FLF-$n$. Finally, we note that the free energy grows linearly with $\nu$ for large value of the fluctuating field. It assures us that the integration in (\ref{eq:FLF_average_sym}) gives finite result for the wide range of observables $\tilde{\mathcal{O}}(\nu)$, which grow asymptotically slower than $e^{-\nu^2}$.

\section{Single-particle quantities}
\label{sec:single}
\begin{figure*}[!t]
\centering
\begin{subfigure}{.5\textwidth}
  \captionsetup{justification=centering}
  \includegraphics[width=1.\linewidth]{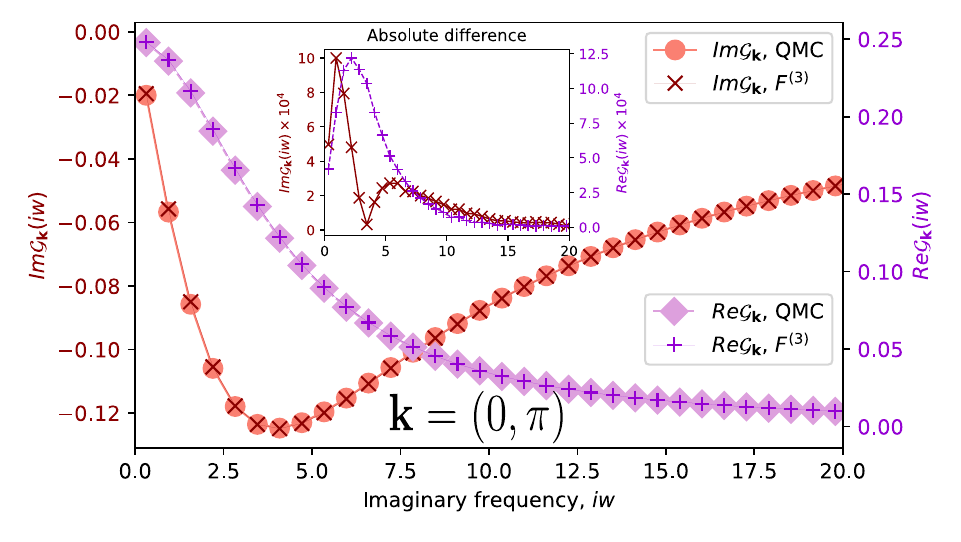}
  \label{fig:sub1}
\end{subfigure}%
\begin{subfigure}{.5\textwidth}
  \captionsetup{justification=centering}
  \includegraphics[width=1.\linewidth]{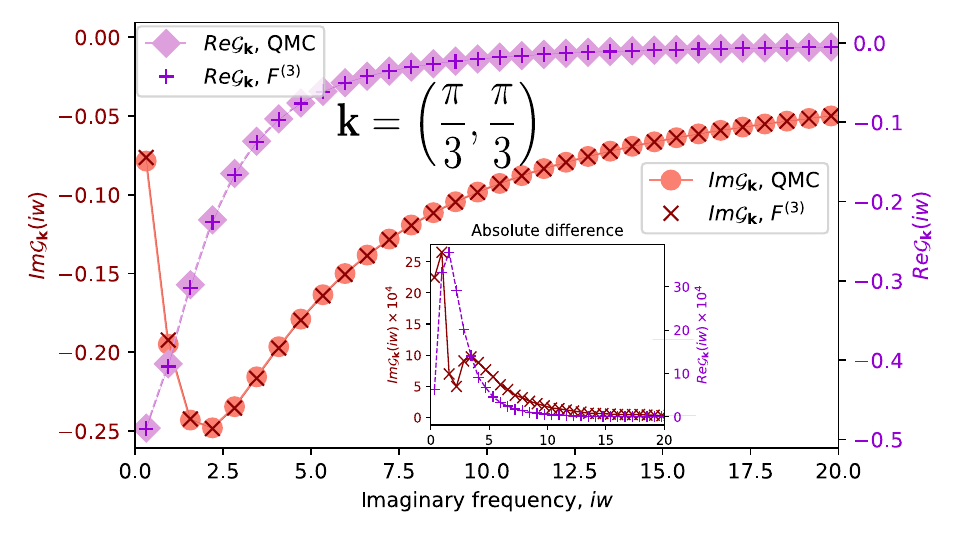}
  \label{fig:sub2}
\end{subfigure}
\caption{\label{fig:Gk} Imaginary and real parts of the Green's functions $\mathcal{G}_{\boldsymbol{k}} (iw)$ in $\boldsymbol{k}$-space near Fermi surface. Comparison of the numerical results obtained within the 3rd order FLF approximation with the QMC results. The inset plot shows the absolute difference between QMC and 3rd order FLF data. The parameters of the Hubbard system are $U/t = 1$, $\beta = 10$, the lattice size is $6 \times 6$.}
\end{figure*}
The FLF ensemble (\ref{eq:FLF_average}) provides us with a powerful tool to investigate the properties of many-particle systems without complex techniques. Here we start with calculation of single-particle quantities.
The temperature Green's function in the frequency domain can be defined as
\begin{equation}
    \mathcal{G}_{\boldsymbol{k}}(iw_n) = \int_0^\beta \mathcal{G}_{\boldsymbol{k}}(\tau)e^{iw_n\tau} d\tau,
    \label{eq: G}
\end{equation}
where
\begin{equation}
    \mathcal{G}_{\boldsymbol{k}} (\tau) = -{\cal Z}^{-1}\int c_{\boldsymbol{k}} (\tau) c_{\boldsymbol{k}}^{\dag} (0) e^{-[\mathcal{S}^0_h+\mathcal{S}_U]} \mathcal{D}\left[c^\dag, c\right].
\end{equation}
Here $\boldsymbol{k}$ denotes a quasimomentum vector, and $iw_n = i{\pi \left(2n+1 \right)}{\beta^{-1}}$ is the $n$th fermionic Matsubara frequency.

According to the FLF approach we can approximate (\ref{eq: G}) with the expectation value over the ensemble of non-interacting systems perturbed by new interaction term $\mathcal{S}_{int}^{\prime}$ (\ref{eq:FLF action}), which is a small parameter for $\lambda = \frac{U}{2}$ as it was shown in the previous section. Since perturbations are already taken into account by (\ref{eq:Z_approx}), we approximate the exact Green's function $\mathcal{G} (iw_n)$ by the bare Green's function $G_k^{\nu} (iw_n)$ averaged over the FLF ensemble
\begin{equation}
    \mathcal{G}_{\boldsymbol{k}} (iw_n) \approx \langle G_{\boldsymbol{k}}^{\nu} (iw_n) \rangle_{FLF} = \av{\frac{1}{iw_n - e_{\boldsymbol{k}}^{\nu}}}_{FLF}.
\label{eq:G_flf}
\end{equation}
Here $e_{\boldsymbol{k}}^{\nu}$ is the energy of the non-interacting system in the external field $\boldsymbol{\nu}$.
\begin{figure*}[!t]
\centering
\begin{subfigure}{.5\textwidth}
  \captionsetup{justification=centering}
  \includegraphics[width=1.\linewidth]{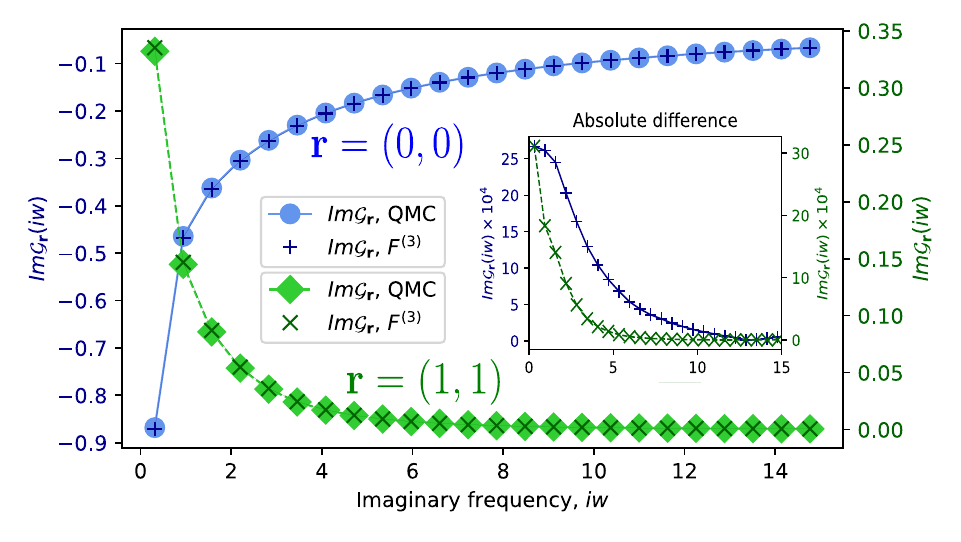}
  \label{fig:sub1}
\end{subfigure}%
\begin{subfigure}{.5\textwidth}
  \captionsetup{justification=centering}
  \includegraphics[width=1.\linewidth]{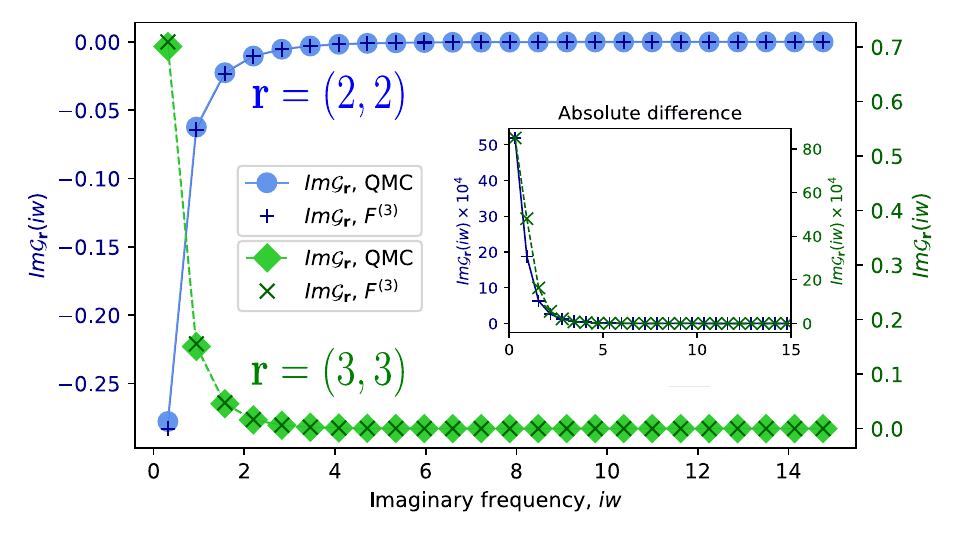}
  \label{fig:sub2}
\end{subfigure}
\caption{\label{fig:Gr} Imaginary parts of the Green's functions $\mathcal{G}_{\boldsymbol{r}} (iw)$ in $\boldsymbol{r}$-space. Comparison of the numerical results obtained within the 3rd order FLF approximation with the QMC results. The inset plot shows the absolute difference between QMC and 3rd order FLF data. The parameters of the Hubbard system are $U/t = 1$, $\beta = 10$, the lattice size is $6 \times 6$.}
\end{figure*}

\begin{figure}
    \centering
    \includegraphics[width=0.9\linewidth]{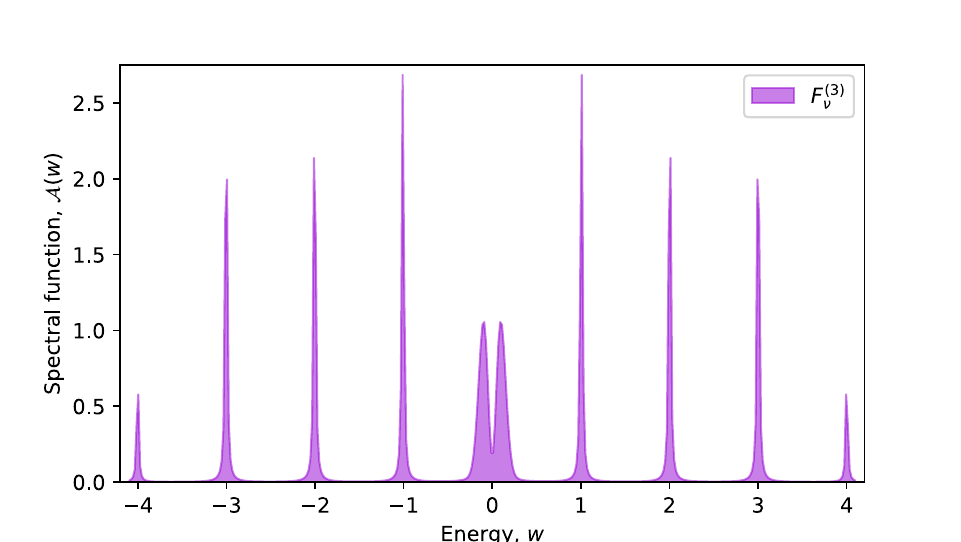}
    \caption{Spectral function $\mathcal{A}(w) = \sum_{\boldsymbol{k}}{\mathcal{A}_{\boldsymbol{k}}(w)}$ calculated in the 3rd order FLF approximation. The parameters of the Hubbard system are $U/t = 1$, $\beta = 10$, the lattice size is $6 \times 6$.}
    \label{fig:A}
\end{figure}

In Fig. \ref{fig:Gk} and Fig. \ref{fig:Gr} we represent the numerical results for the Green's function $\mathcal{G}(iw)$ in $\boldsymbol{k}$-space and in $\boldsymbol{r}$-space respectively, calculated in the 3rd order FLF approximation. The comparison with the numerically exact Quantum Monte-Carlo (QMC) results shows that the FLF Green's functions perfectly matches with numerically exact ones at moderately low temperature $\beta = 10$. This allows us to obtain reliable experimentally observable quantities without complex computational techniques. For instance, to obtain the momentum-resolved spectral function $\mathcal{A}_{\boldsymbol{k}} (w) $ in the QMC approach one would have to deal with analytic continuation of the QMC data. To do the same in the FLF scheme we calculate the local Green's function in the real frequencies domain $\mathcal{G}_{\boldsymbol{k}} \left( w\right)$ by the formal substitution $iw_n \rightarrow w + i0$:
\begin{equation}
    \mathcal{A}_{\boldsymbol{k}} (w) = -\frac{1}{\pi}\Im{\mathcal{G}_{\boldsymbol{k}} (w)} \approx -\frac{1}{\pi} \Im{ \left\langle \frac{1}{w - e_{\boldsymbol{k}}^{\nu}}  \right\rangle_{FLF}}.
\label{eq:A}
\end{equation}
The numerical results for the spectral function $\mathcal{A}(w) = \sum_{\boldsymbol{k}} \mathcal{A}_{\boldsymbol{k}}(w)$ in the FLF approximation are shown in the Fig. \ref{fig:A}. We observe a pseudogap at the Fermi level $w=0$, which indicates violation of Fermi-liquid behavior due to the perfect nesting at half-filling \cite{PhysRevX.11.011058}. We note however that, in our calculations, the pseudogap does not close for small $\beta$ which is caused by the breakdown of collective modes at high temperatures. We also comment on this in the Section \ref{sec:conclusion}.

\section{Susceptibility in the antiferromagnetic channel}
\label{sec:susceptibility}

In order to further explore the power of the FLF ensemble we consider susceptibility $\chi$ in the spin-channel. As we already mentioned, the half-filled Hubbard model possesses fluctuations in the AF channel. Mean-field-like schemes predict unphysical transition to the AF-ordered state at finite temperature, which manifests as the infinite susceptibility to the external AF magnetic field $\boldsymbol{h}$. The possible cause of this effect is that MF schemes freeze all the fluctuations, which actually break the ordered states. If this assumption is valid, taking into account AF fluctuations in the FLF approach would cure this artifact.

The magnetic susceptibility can be defined as
\begin{equation}
\label{eq:chi_def}
    \chi = \left.\frac{1}{\beta N} \partial^2_{\boldsymbol{h}}\ln Z\right|_{\boldsymbol{h}\to 0}.
\end{equation}

Using the expression (\ref{eq:FLF_transf simplified}) for the partition function of the system and the definition of average over the FLF ensemble (\ref{eq:FLF_average}) we obtain:
\begin{equation}
    \chi = \frac{\beta N}{3\lambda^2}\langle\nu^2\rangle_{FLF} - \frac{1}{\lambda}.
\end{equation}
The $1/3$ coefficient appeared due to the spatial isotropy.

\begin{figure}
    \centering
    \includegraphics[width=0.98\linewidth]{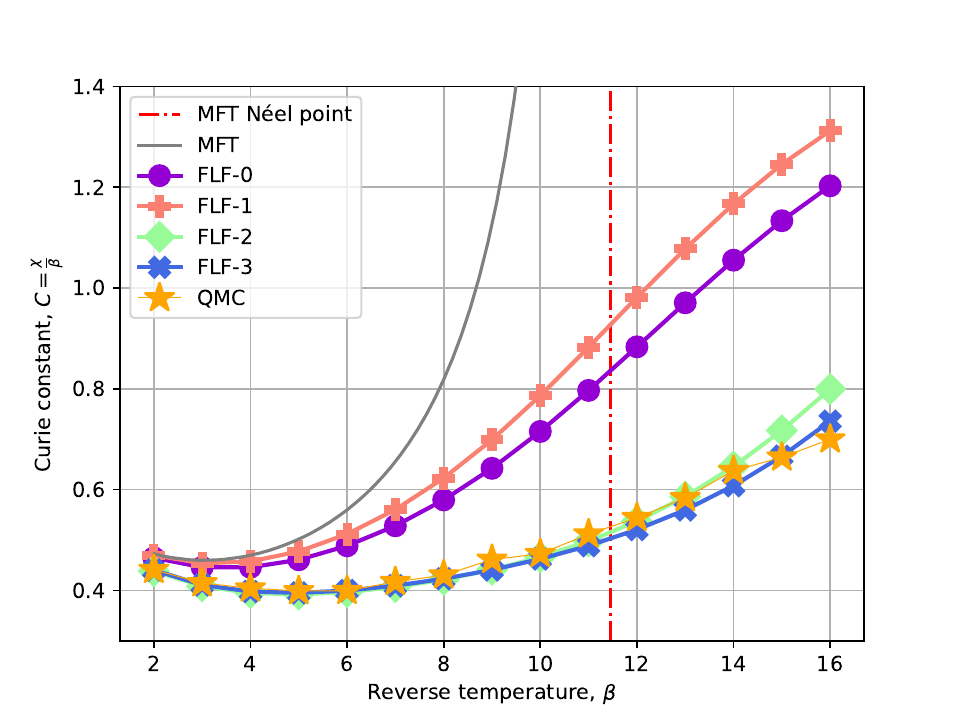}
    \caption{Curie constant $C=\chi/\beta$ as a function of the inverse temperature $\beta$. Comparison of the numerical results obtained within different orders of FLF approximation with the MFT and QMC results. The parameters of the Hubbard system are $U/t = 1$, the lattice size is $6\times 6$.}
    \label{fig:Curie}
\end{figure}

The numerical results are presented in terms of the Curie constant $C=\chi/\beta$, and are shown in the Fig. \ref{fig:Curie}. We considered the 2D Hubbard lattice with $6\times 6$ sites at $U = t = 1$. We compared the results taking into account from $0$th to $3$rd order corrections with the mean field theory (MFT) and with the QMC results. One can see that FLF-3, MFT and QMC methods all lead to coinciding results for comparatively high temperatures $\beta < 4$. At $\beta\approx 11.5$ though the MFT predicts N\'eel transition to the ordered phase, which is unphysical. We also note that even more sophisticated DMFT method, which accounts for local effects exactly, predicts such a transition, though at slightly lower temperatures (see i.e. \cite{PhysRevX.11.011058}, \cite{RSL-2020}). It indicates two important points. The first point is that the Hubbard lattice possesses correlation effects even at low interaction regime. And the second point is that the fluctuations responsible for this are non-local ones. FLF scheme describes the physics around this temperature qualitatively correct. Quantitative correctness of the FLF results can be concluded from the comparison with the QMC data. Though statistical noise leads to some errorbar of the QMC points, one can see that the FLF-3 results are nearly within the margin of error of the numerically exact QMC method.

\section{Total energy}
\label{sec:E_av}
In Section \ref{sec:Perturbation} we saw, that the FLF method allows to calculate the partition function perturbatively. Once this is done, one can find all the thermodynamic quantities of the system. In this Section, we consider the total energy per site of the Hubbard lattice as an example. By definition,
\begin{figure}
    \centering
    \includegraphics[width=0.98\linewidth]{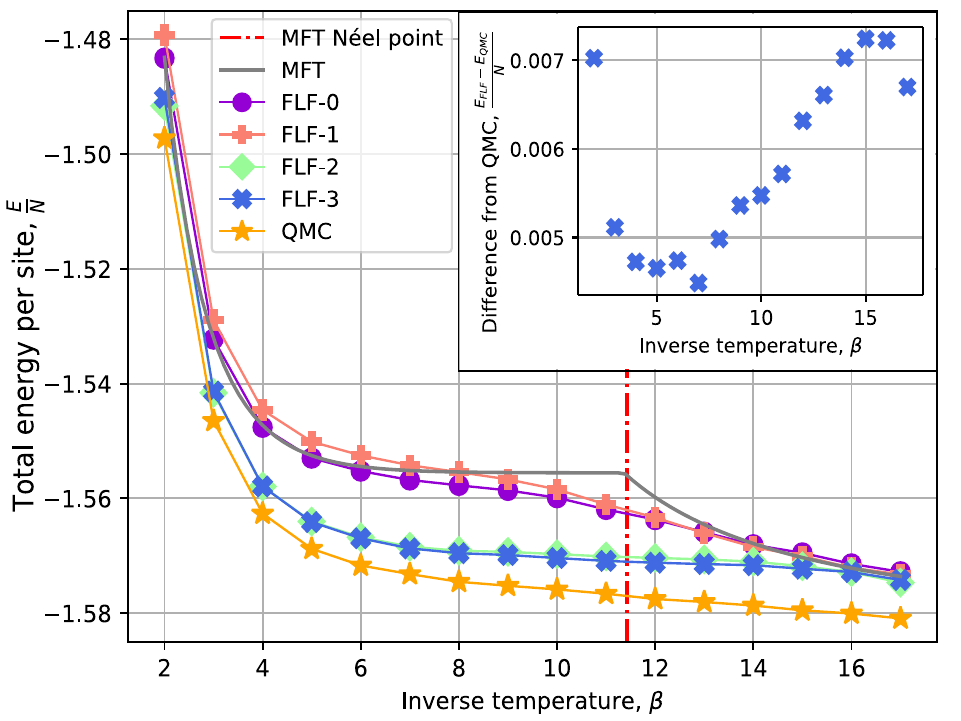}
    \caption{Total energy per site $E/N$ as a function of the inverse temperature $\beta$. Comparison of the numerical results obtained within different orders of FLF approximation are presented. The total energy per site is compared with the MFT and QMC results. The parameters of the Hubbard system are $U/t=1$, the lattice size is $6 \times 6$.}
    \label{fig:E_total}
\end{figure}

\begin{equation}
\frac{E}{N} = -\frac{1}{N} \partial_\beta \ln{Z}.
\label{E_tot}
\end{equation}
Here, one can use the finite difference method to calculate the derivative numerically. In the Fig. (\ref{fig:E_total}), we present the results obtained within different orders of the FLF perturbative approach in comparison with the MFT and QMC results. At first, it is clearly seen that all FLF orders show more accurate results that the MFT ones, which significantly differ from the QMC data even at the remarkably high temperatures $\beta \gtrsim 3$. Furthermore, the MFT curve exhibits a kink corresponding to the transition to the ordered state, which is absent in the numerically exact QMC data. The FLF approach cures this artifact at all perturbation orders, leading to qualitatively correct results. One can see that the FLF-0 and FLF-1 curves are quantitatively inaccurate. Nevertheless, higher FLF orders give much more closer to the QMC results. The inset plot shows the quantitative difference between the FLF-3 and QMC methods which is largest at high temperatures $\beta < 3$ and after the point of the MF phase transition $\beta \approx 11.5$. It can be understood as follows. In the high temperature regime, the collective antiferromagnetic mode accounted for by the FLF approach is absent, so the theory becomes inapplicable. And after the N\'eel transition point, the fluctuations become too strong, so that the FLF approach fails to accurately reproduce them. These difficulties can be overcame, though, in further developments of the proposed FLF scheme. The relevant detailed discussion is presented in Section \ref{sec:conclusion}.

\section{Cluster extension}
\label{sec:cluster}
Assuming that fluctuations belong to a single mode we limit the available range of systems with small clusters. However, the FLF approach could still be applied to larger systems in the spirit similar to cluster DMFT extensions. In this section we develop cellular FLF (c-FLF) method in the CDMFT-like manner \cite{DCArev}.

We start with the definition of the self-energy
\begin{equation}
    \Sigma_{\boldsymbol{k}}(z_n) = z - e_{\boldsymbol{k}} - \mathcal{G}_{\boldsymbol{k}}(z_n)^{-1},
\label{eq:Dyson self-energy}
\end{equation}
where $z$ is a real $z_n = w_n + i0$ or imaginary frequency $z_n = iw_n$. We make use of the coarse-graining and decompose the momentum vector
in two parts $\boldsymbol{k} = \tilde{\boldsymbol{k}} + \delta \boldsymbol{k}$, where $\tilde{\boldsymbol{k}}$ is the coarsened vector defined on the finite size cluster and $\delta \boldsymbol{k}$ is the residual vector. Using (\ref{eq:G_flf}) as the approximation for the cluster Green's function we obtain the cluster self-energy
\begin{equation}
    \Sigma_{\tilde{_{\boldsymbol{k}}}}(z) = z - e_{\tilde{_{\boldsymbol{k}}}} -  \left\langle \left(z - e_{\tilde{_{\boldsymbol{k}}}}^{\nu} \right)^{-1}  \right\rangle^{-1}_{FLF}.
\end{equation}
Assuming now that the essential interaction physics is captured by the cluster self-energy $\Sigma_{\boldsymbol{k}}(z) \approx \Sigma_{\tilde{_{\boldsymbol{k}}}}(z)$ we go back to (\ref{eq:Dyson self-energy}) and finally obtain the lattice Green's function
\begin{equation}
   \mathcal{G}_{\boldsymbol{k}}(z) = \frac{1}{z - e_{\boldsymbol{k}} - \Sigma_{\boldsymbol{\tilde{k}}}(z)}.
\end{equation}
This way we can describe a large system with many fluctuating modes in terms of a finite-size cluster with a single mode fluctuations, and thus overcome this limitation while preserving the low computational cost of the numerical procedure.

\begin{figure}[!h]
\centering
\includegraphics[width=1.\linewidth]{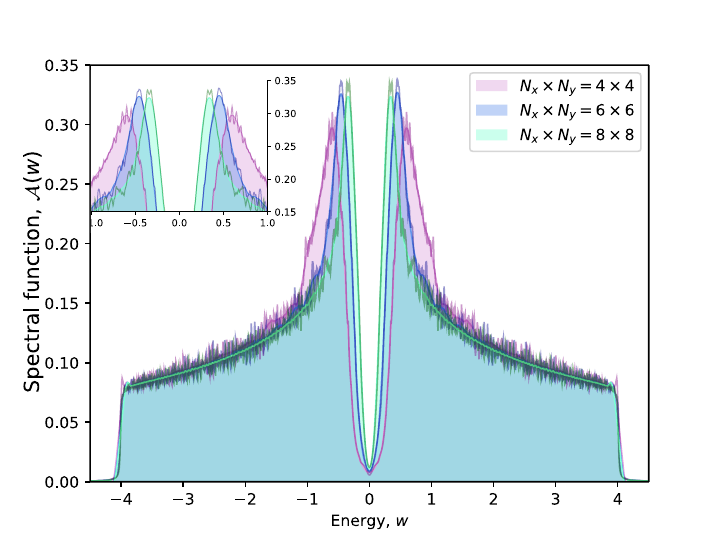}
\caption{Local density of states Spectral function $\mathcal{A}(w)$ calculated within the c-FLF approach with clusters of different sizes $N_x \times N_y$. The Hubbard system parameters are $U/t = 1$, $\beta = 10$.}
\label{fig:Cluster}
\end{figure}

In Fig. \ref{fig:Cluster} we show the results for the spectral function $\mathcal{A}(w)$ for infinite lattice calculated in the c-FLF approach with clusters of different sizes. All the considered cluster sizes lead to almost the same results indicating the convergence. In the Fig. \ref{fig:Cluster} we may observe a pseudogap, which can be due to two reasons. The first one is the finite size of clusters. One can see that the pseudogap indeed reduces with the growth of the cluster size. The second one may take origin in the approximation of the exact Green's function $\mathcal{G}_{\boldsymbol{k}}(z)$ by the ensemble of non-interacting ones (\ref{eq:G_flf}). It seems reasonable that this artifact could be suppressed by a more sophisticated approximation for $\mathcal{G}_{\boldsymbol{k}}(z)$,  although this point in the Hubbard phase diagram is very close to the metal-insulator crossover calculated by the diagrammatic Monte Carlo \cite{Kozik2020}.

\section{Convergence issues and relation to Landau free energy}
\label{sec:Landau}
\begin{figure}[!h]
\centering
\includegraphics[width=1.\linewidth]{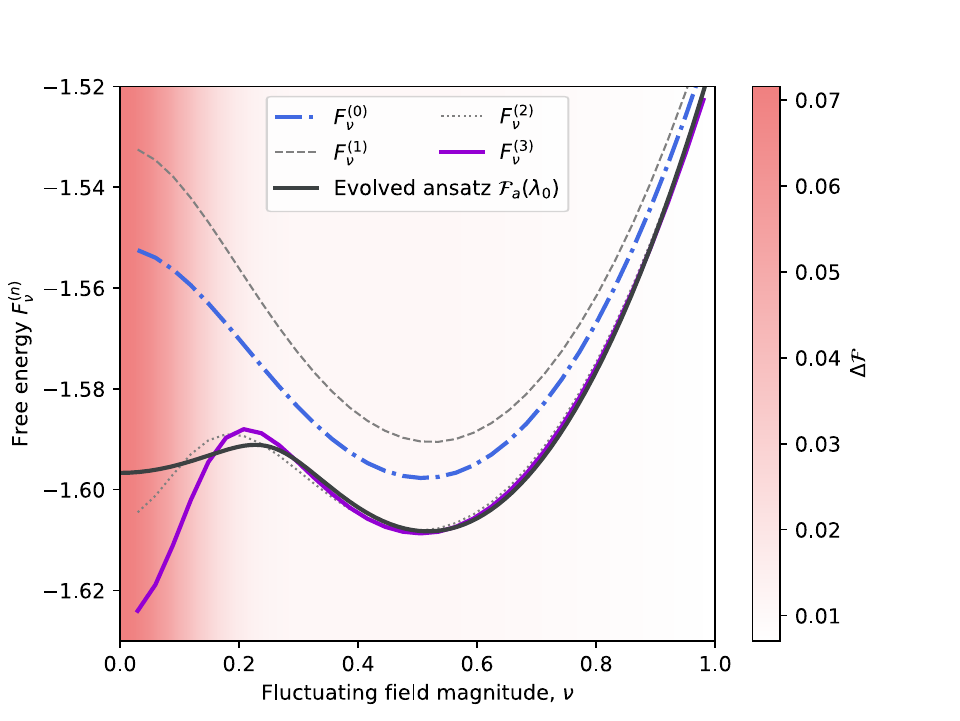}
\caption{Comparison of the 0th $F^{(0)}$ and 3rd $F^{(3)}$ order FLF free energies with the regularized ansatz function $\mathcal{F}_{\boldsymbol{a}}$. Color map indicates the difference $\Delta F$ between the 0th and the 3rd order FLF free energies. The Hubbard system parameters are $U/t=2$, $\beta=10$, the lattice size is $4\times 4$.}
\label{fig:Ansatz}
\end{figure}

In the previous sections we considered the numerical results obtained in the FLF approach based on the perturbation series for free energy (\ref{eq:F_series}). We noticed that this rather simple-to-implement method leads us to good accuracy for comparatively high temperatures ($\beta \ge 10$) and for weakly-medium correlated systems $U/t < 2$. In this section we address the question, to which extent the proposed FLF perturbation scheme is convergent. To this aim, let's consider the applicability of the FLF scheme to systems at lower temperatures and/or in stronger correlated regime. In Fig. \ref{fig:Ansatz} we demonstrate the 0th and the 3rd order FLF free energy approximation as the function of fluctuating field magnitude $\nu$ for $U/t=2$. One can see that the discrepancy between these two curves for $\nu\to 0$, seeming harmless for $U/t=1$ (see Fig. \ref{fig:F_series}), seeds the divergence for $U/t=2$ in this region. It makes us unable to use the FLF perturbative approach as is for this regime. In this section we develop the regularization procedure based on the idea of Landau free energy flow which would allow us to extend the range of the method applicability.

Let us consider the half-filled Hubbard cluster in the view of phenomenological Landau theory of phase transitions. This way, we can describe the system in terms of the order parameter, which is the collective AF spin $S^l$ (\ref{eq:total spin}) in our case. Then we can represent the exact partition function (\ref{eq:Z_exact}) as the double integral
\begin{equation}
\label{eq:LFE 0}
    Z = \int d^3 \tilde{s} \int \displaylimits_{\boldsymbol{s}[c^\dag,c] = \tilde{\boldsymbol{s}}} e^{- \mathcal{S}_{h}[c^\dag,c;\boldsymbol{s}]}  \mathcal{D} \left[ c^{\dagger}, c\right]
\end{equation}
of $e^{-\mathcal{S}_h}$ over Grassman variables $c^\dag,c$ at fixed values of the order parameter $\tilde{\boldsymbol{s}}$. It leads us to the definition of the Landau free energy $F_L(\tilde{\boldsymbol{s}})$:
\begin{equation}
\label{eq:F_Landau}
    Z = \int  e^{- \beta N F_L \left(  \tilde{\boldsymbol{s}}\right)} d^3 \tilde{s}.
\end{equation}
Reformulating the FLF transformation (\ref{eq:FLF_transf}) in this spirit we obtain
\begin{equation}
\label{eq:LFE-FLF}
   Z = \iint  e^{- \beta N F_{L} \left(  \tilde{\boldsymbol{s}}\right)} e^{- \frac{\beta N \lambda}{2} \left(\tilde{\boldsymbol{s}} - \tilde{\boldsymbol{s}}' \right)^2}  d^3 \tilde{s} d^3 \tilde{s}',
\end{equation}
where $\tilde{\boldsymbol{s}}' = \boldsymbol{\nu\lambda}$, and we omitted the dependence on $\boldsymbol{h}$ for the sake of shortness. Integrating this over the Grassman variables at some given value of $\lambda$, we can rewrite this as
\begin{equation}
\label{eq:LFE-FLF}
   Z = \int  e^{- \beta N \mathcal{F}(\tilde{\boldsymbol{s}},\lambda)} d^3 \tilde{s},
\end{equation}
where $\mathcal{F}(\tilde{\boldsymbol{s}},\lambda)$ is a double variable function. At this point one may notice that expressions (\ref{eq:F_Landau})--(\ref{eq:LFE-FLF}) formally describe the solution of the heat flow Cauchy problem
\begin{equation}
\begin{cases}
      \left(\partial_{\lambda^{-1}} - \Delta_{\tilde{s}}\right) e^{-\beta N \mathcal{F}(\tilde{\boldsymbol{s}},\lambda)} = 0,\\
      \left.e^{-\beta N \mathcal{F}(\tilde{\boldsymbol{s}},\lambda)}\right|_{\lambda^{-1}\to 0} = e^{-\beta N F_L(\tilde{\boldsymbol{s}})},
\end{cases}
\label{Backward cauchy problem}
\end{equation}
obtained by a convolution (\ref{eq:LFE-FLF}) of the fundamental solution $\text{exp}\left(-\beta N\lambda\tilde{\boldsymbol{s}}^2/2\right)$ with the initial condition of (\ref{Backward cauchy problem}). Here $\lambda^{-1}$ plays the role of time variable, and $(2\beta N)^{-1}$ can be considered as the thermal conductivity.

This observation allows us to derive the regularization procedure based on the flow of free energy. Our aim is to find some function $\mathcal{F}(\tilde{\boldsymbol{s}},\lambda)$ that best suits the 3rd order FLF free energy $F^{(3)}$, but has a more regular behavior at $\nu\to 0$. The backward Cauchy problem (\ref{Backward cauchy problem}) is ill-conditioned, so we will not solve it directly. Instead, we set the initial function to be an ansatz $\mathcal{F}_{\boldsymbol{a}}(\tilde{\boldsymbol{s}})$ depending on vector of parameters $\boldsymbol{a}$, and let it evolve through the convolution (\ref{eq:F_Landau}). Since the physically meaningful range of the order parameter is $\tilde{s} \le 1$, we impose an additional condition
\begin{equation}
    \left.e^{-\beta N \mathcal{F}(\tilde{\boldsymbol{S}},\lambda)}\right|_{\tilde{S}>1} = 0.
\end{equation}
The particular value of $\boldsymbol{a}$ would be the one at which $\mathcal{F}_{\boldsymbol{a}}(\tilde{\boldsymbol{S}},\lambda_0)$ is the closest to the 3rd order FLF Landau free energy $F^{(3)}(\tilde{\boldsymbol{S}},\lambda_0)$ at $\lambda_0$. As a measure of similarity, we use the loss function
\begin{equation}
    \mathcal{L}_{\alpha}  = \int \frac{\left|\mathcal{F}_{\boldsymbol{a}}(\tilde{\boldsymbol{s}},\lambda_0)  -  F^{(3)}(\tilde{\boldsymbol{s}},\lambda_0)\right|^2}{\alpha (\Delta F)^2 + 1} d^3\tilde{s} \rightarrow min,
\label{Cost function}
\end{equation}
where 
\begin{equation}
    \Delta F = \max\limits_{\tilde{s} > \tilde{s}^{\prime}} \left|F^{(3)} (\tilde{s}^{\prime}) - F^{0} (\tilde{s}^{\prime})\right|
\end{equation}
is the disconvergence between the 0th and the 3rd order FLF free energy, and $\alpha$ is the regularization constant. The denominator of this loss function (\ref{Cost function}) is constructed so that it reduces the contribution of the dominant disconvergence region. In particular, we choose the value of the $\alpha$ parameter such that $\alpha (\Delta F)^2 \approx 1$. Once the values of $\boldsymbol{a}$ are found, we will be able to use $\mathcal{F}_{\boldsymbol{a}}$ the same way as the FLF free energy to calculate observables (\ref{eq:FLF_average}).

In the particular case of negligible external magnetic field $\boldsymbol{h}\to 0$ the Hubbard system possesses spherical symmetry, and we can use the finite even-order polynomial ansatz
\begin{equation}
    \mathcal{F}_{\bold{a}}(\tilde{\boldsymbol{s}}) = a_0 + a_2 \tilde{s} ^2 + ... + a_{2p} \tilde{s}^{2p},\quad p\in \mathbb{N}.
\label{Ansatz}
\end{equation}

We present the results for the $\mathcal{F}_{\boldsymbol{a}}(\nu) = \mathcal{F}_{\boldsymbol{a}}(\tilde{\boldsymbol{s}}/\lambda)$ obtained for $4\times 4$ Hubbard lattice at $U/t=2$ and $\beta=10$ in Fig. \ref{fig:Ansatz}. In this case, we set $\alpha = 10^6$, and we limited ourselves with the order of polynomial $p=5$. Nevertheless, we should mention that as the temperature decreases, the behavior of the Landau free energy $F_L$ becomes more complex, and the order of the polynomial has to be increased. We also note that the problem posed is not convex, and therefore has many local minima. To estimate the global one, we used the \textit{SHGO} method \cite{SHGO} implemented in \textit{SciPy} python package \cite{SciPy}.
\begin{figure*}
\centering
\begin{subfigure}{0.42\textwidth}
  \captionsetup{justification=centering}

  \includegraphics[width=1.\linewidth]{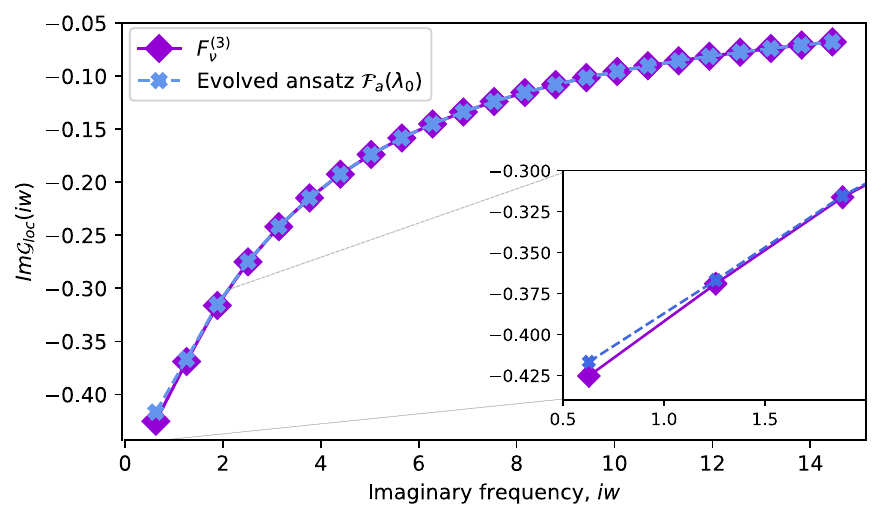}
  \caption{}
  \label{fig:anz_greens}
\end{subfigure}
\begin{subfigure}{0.46\textwidth}
  \captionsetup{justification=centering}
  \includegraphics[width=1.\linewidth]{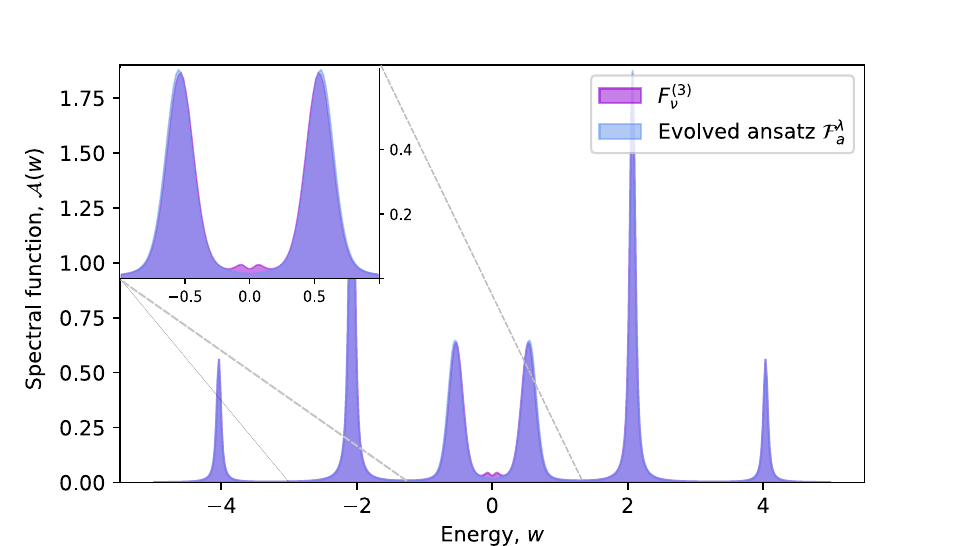}
  \caption{}
  \label{fig:anz_dos}
\end{subfigure}%
\caption{Comparison of the numerical results obtained within the 3rd order FLF approximation with the regularized approximation. Imaginary part of the local Green's function (a) and the spectral function (b). The parameters of the Hubbard system are $U/t=2$, $\beta=10$, the lattice size is $4 \times 4$.}
\label{Ansatz greens function}
\end{figure*}
One can see that the evolved ansatz function almost coincides with the 3rd order FLF free energy $F^{(3)}$ for $\nu>0.5$, and the divergence for $\nu\to 0$ is significantly smoother.

As an example of how this regularization manifests in observables, we provide the numerical results for the local Green's function and spectral function $\mathcal{A}(w)$ in Fig. \ref{Ansatz greens function}. Comparison of the pure 3rd order FLF approximation with the regularized approximation demonstrates that the second one slightly modifies local Green's function for small imaginary frequencies $iw$ (see Fig. \ref{fig:anz_greens}). What is more noteworthy, regularization approach leads to the appearance of peaks near the Fermi level in the spectral function in Fig. \ref{fig:anz_dos}.

\section{Conclusion and outlook}
\label{sec:conclusion}
In this work we derived the extended version of the Fluctuating Local Field method using the perturbation theory approach. We considered the half-filled 2D Hubbard lattices, which are known to possess collective spin fluctuations, as the particular example. The FLF scheme, originally based on the exact transformation of the partition function, introduces an auxiliary classical fluctuating field coupled to the collective order parameter of the system in the leading fluctuation channel. For the system under consideration it leads to the emergence of an effective long-range interaction $\lambda$ competing with the local interaction $U$. We show that for $\lambda_0=U/2$ these two interactions form a small parameter, enabling us to construct a regular series expansion w.r.t. $U$.

In order to investigate the efficiency of the proposed FLF perturbation scheme we calculated several physical quantities, namely Green's function, spectral function, magnetic susceptibility, and total energy. Comparison with much more sophisticated DMFT method, and with the numerically exact QMC method shows that the FLF perturbation approach leads to good accuracy preserving the comparative simplicity of the implementation.

To obtain physically consistent results, one should keep in mind the range of applicability of the proposed FLF scheme. First of all, as the series expansion is constructed w.r.t. the powers of $U$, it better suits weak-coupling regime. Nevertheless, the FLF perturbation series allows us to go beyond the limitations of usual perturbation schemes and leads to good accuracy even for $U/t=1$. And the proposed regularization procedure, based on the flow of the Landau free energy, even extends the range of applicability to $U/t \lesssim 2$, which already falls within the range of moderately correlated regime. Handling lattices with larger $U/t$ would require an FLF scheme constructed around DMFT, which will be in the scope of our future works.

Secondly, number of the lattice sizes should be large enough to form a collective degree of freedom. At the same time, the lattice should not be too large, as we expect that fluctuations belong to a single mode. Similar argumentation holds for the temperature range. Formally it is reflected in a diagram hierarchy in powers of $(\beta N)^{-1}$. Nevertheless, we show that the limitation on the size of the lattice can be overcome using the cluster extension of the FLF perturbation scheme.

Regardless of the fact that we focused on the single mode fluctuations in the spin channel in this work, the flexibility of the FLF approach actually allows one to describe the physics of the systems with other types of fluctuations, both in single mode and multi mode regimes. Describing the physics of systems with competing fluctuation channels will be in the scope of our future work, as well as the FLF perturbation extension around DMFT.

\section*{Acknowledgements}

Ya.S.L., S.D.S., and A.N.R. acknowledge the support from the Russian Quantum Technologies Roadmap.

\appendix*
\onecolumngrid
\section{Third order corrections}
This section is dedicated to the derivation of the third order corrections to the free energy. We start from (\ref{eq: Taylor expansion}) and write explicitly the third term of partition function expansion $\delta\mathcal{Z}^3_\nu$ coming from $e^{-\mathcal{S}'_{int}}$:
\begin{equation}
    \delta\mathcal{Z}^3_\nu = -\frac{1}{6}\int\left[U^3\theta\theta\theta + \frac{3\beta NU^2\lambda}{2} S^lS^l\theta\theta + \frac{3(\beta N)^2U\lambda^2}{4} S^lS^l S^mS^m \theta + \frac{(\beta N\lambda)^3}{8}S^lS^lS^mS^mS^pS^p\right]e^{-\mathcal{S}_0^\nu}\mathcal{D}\left[c^\dag, c\right].
\end{equation}
Using the trick (\ref{eq:derivative_trick}), we can rewrite this in the following form:
\begin{equation}
    \delta\mathcal{Z}^3_\nu = \left[-\frac{U^3}{6}\langle \theta\theta\theta\rangle_0 - \frac{\lambda U^2}{4\beta N}\Delta_\nu\langle \theta\theta\rangle_0 - \frac{\lambda^2 U}{8(\beta N)^2} \Delta_\nu\Delta_\nu \langle \theta\rangle_0 - \frac{\lambda^3}{48(\beta N)^3}\Delta_\nu\Delta_\nu\Delta_\nu\right]\mathcal{Z}_\nu^0.
\end{equation}
Finally, for the third order correction to the free energy $\delta F^3_\nu$ we have
\begin{align}
    \delta F^{3}_\nu &= \frac{U^3}{6}\Theta_3 + \frac{\lambda U^2}{2}\left(-\Theta_2'(F_\nu^0)' + \Theta_1^{'2}\right) + \frac{\lambda U^2}{2\beta N}\left(\frac{\Theta_2'}{\nu} + \frac{\Theta_2^{''}}{2}\right) +\\
    \nonumber
    &+\frac{\lambda^2U}{2}\left(\Theta_1^{''}(F_\nu^0)^{'2} + 2\Theta_1'(F_\nu^0)'(F_\nu^0)^{''}\right) +\\
    \nonumber
    &+\frac{\lambda^2U}{2\beta N}\left(- \frac{2\Theta_1'(F_\nu^0)^{''}}{\nu} - \Theta_1^{''}(F_\nu^0)^{''} + \frac{2\Theta_1'(F_\nu^0)'}{\nu^2}-\frac{2\Theta_1^{''}(F_\nu^0)'}{\nu} - \Theta_1^{'''}(F_\nu^0)' - \Theta_1'(F_\nu^0)^{'''}\right)+\\
    \nonumber
    &+\frac{\lambda^2U}{2(\beta N)^2}\left(\frac{\Theta_1^{(IV)}}{4}+\frac{\Theta_1^{'''}}{\nu}\right)+\\
    \nonumber
    &+\frac{\lambda^3}{2}\left((F_\nu^0)^{'2}(F_\nu^0)^{''2}+\frac{(F_\nu^0)^{'3}(F_\nu^0)^{'''}}{3}\right)+\\
    \nonumber
    &+\frac{\lambda^3}{2\beta N}\left(-\frac{(F_\nu^0)^{''3}}{3} - \frac{2(F_\nu^0)'(F_\nu^0)^{''2}}{\nu} - \frac{2(F_\nu^0)^{'3}}{3\nu^3}+ \frac{(F_\nu^0)^{'2}(F_\nu^0)^{'''}}{\nu}+\right.\\
    \nonumber
    &\left.+\frac{2(F_\nu^0)^{'2}(F_\nu^0)^{''}}{\nu^2} - \frac{2(F_\nu^0)^{'2}(F_\nu^0)^{'''}}{\nu}- \frac{(F_\nu^0)^{'2}(F_\nu^0)^{(IV)}}{2} - 2(F_\nu^0)'(F_\nu^0)^{''}(F_\nu^0)^{'''}\right)+\\
    \nonumber
    &+\frac{\lambda^3}{2(\beta N)^2}\left( \frac{(F_\nu^0)'(F_\nu^0)^{(V)}}{4} + \frac{5}{12}(F_\nu^0)^{'''2}-\frac{(F_\nu^0)'(F_\nu^0)^{'''}}{\nu^2}+\frac{2(F_\nu^0)^{''}(F_\nu^0)^{'''}}{\nu}+\right.\\
    \nonumber
    &+\left.\frac{(F_\nu^0)'(F_\nu^0)^{(IV)}}{\nu}+ \frac{(F_\nu^0)^{''}(F_\nu^0)^{(IV)}}{2} \right)+\\
    \nonumber
    &+\frac{\lambda^3}{2(\beta N)^3}\left(-\frac{(F_\nu^0)^{(V)}}{4\nu}- \frac{(F_\nu^0)^{(VI)}}{24}\right).
\end{align}
Here $\beta N \Theta_3 = 2\av{\theta}_0\av{\theta}_0\av{\theta}_0 - 3\av{\theta}_0\av{\theta\theta}_0 + \av{\theta\theta\theta}_0$, and the terms are grouped according to their extensity/intensity.

\twocolumngrid
\bibliography{Ref}
\end{document}